\documentclass{article}

    \PassOptionsToPackage{numbers, compress}{natbib}

    \usepackage[final]{neurips_2022}

\usepackage{caption} 
\usepackage{graphicx} 
\usepackage[utf8]{inputenc} 
\usepackage[T1]{fontenc}    
\usepackage{hyperref}       
\usepackage{url}            
\usepackage{booktabs}       
\usepackage{amsfonts}       
\usepackage{nicefrac}       
\usepackage{microtype}      
\usepackage{xcolor}         

\title{Optimized Global Perturbation Attacks For Brain Tumour ROI Extraction From Binary Classification Models}

\usepackage{authblk}
\author[1,3]{Sajith M. Rajapaksa}
\author[1,2,3,4,5,6,7,*]{Farzad Khalvati}
\affil[1]{Neurosciences and Mental Health, The Hospital for Sick Children, Toronto, Canada}
\affil[2]{Department of Diagnostic Imaging, The Hospital for Sick Children}
\affil[3]{Institute of Medical Science, University of Toronto}
\affil[4]{Department of Medical Imaging, University of Toronto}
\affil[5]{Department of Mechanical and Industrial Engineering, University of Toronto}
\affil[6]{Department of Computer Science, University of Toronto}
\affil[7]{Vector Institute, Toronto, Canada}

\affil[ ]{\textit {sajith.rajapaksa@mail.utoronto.ca, farzad.khalvati@utoronto.ca*}}

\begin{document}

\bibliographystyle{IEEEtran.bst}
\maketitle

\begin{abstract}
Deep learning techniques have greatly benefited computer-aided diagnostic systems. However, unlike other fields, in medical imaging acquiring large fine-grained annotated datasets such as 3D tumour segmentation is challenging due to the high cost of manual annotation and privacy regulations. This has given interest to weakly-supervise methods to utilize the weakly labelled data for tumour segmentation. In this work, we propose a weakly supervised approach to obtain regions of interest using binary class labels. Furthermore, we propose a novel objective function to train the generator model based on a pretrained binary classification model. Finally, we apply our method to the brain tumour segmentation problem in MRI. 
\end{abstract}

\section{Introduction}

Through learning binary classification, the convolutional neural network (CNN) can also learn the regions of interest to focus on \cite{niu2021review}. By perturbing regions in the input image and comparing the effect on the classification model to the non-perturbed classification, we can understand the model’s reliance on a particular region, which may represent the region of interest (ROI) such as tumour \cite{ribeiro2016should}. Rajapaksa S. and Khalvati F. proposed an iterative approach to obtain segmentation masks by perturbing segmented superpixels \cite{rajapaksa2021localized}. The authors also showed that an optimized perturbation obtains the most information from the pretrained classification model for a given image. However, since the proposed work is based on an iterative approach, it can consume a large number of computational resources. In this work, we significantly improve upon \cite{rajapaksa2021localized} to remove the iterative perturbation by introducing optimized global perturbation mask generation, which is an iteration-less approach to generate a map indicating which regions are more relevant to the classification task (Relevance Maps). We apply the proposed method to brain tumour segmentation in MRI and show that our method consumes less resources and outperforms the previous methods.

\section{Methods}

Our proposed algorithm can be broken down into two parts. First whole image perturbation generation, where we utilize a pretrained classification model to generate a perturbation mask that is the most effective in changing the class probability. Second, the Relevance Map generation step. In this step, we create superpixel maps using the input images and assign the value of the perturbation mask to generate the ranked regions. This allows us to use the top-ranking regions as the segmentation mask.  

\subsection{Whole Image Perturbation Generation}
In this step, our goal is to generate a perturbation mask that will have the greatest effect on classification probability but have the least amount of change to the original input image. To accomplish this, we utilize a GAN-like \cite{goodfellow2014generative} structure as our generative model. First, the discriminator was replaced with a pretrained 3D tumour type classification model. Then, the generator is trained with our novel objective function to produce the optimal perturbation mask for the given input.  

\subsubsection{Objective Function}
 
We introduce a novel objective function, modified perturbation loss, which maximizes the difference between the classification probability (generated by the pretrained Resent) of the perturbed image and the original image combined with L1-loss and Indecisive penalty. For each batch, the loss would be calculated by applying (multiplying) the generated perturbations mask onto the original MRI and generating a classification using the pretrained Resent. Then we calculate the difference between the predictions of perturbed and non-perturbed images to determine the effectiveness of that generated perturbation. In the following, we explain each component of the loss function in detail.

\textbf{Perturbation Loss}
$\frac{1}{n} \sum_{i=1}^{n}log(\frac{1}{|y_{p} - y_{np}|})$
Where p is perturbed and np is non\_perturbed, and y is the classification prediction. The goal of the perturbation loss penalty is to encourage the generator model to confuse the pretrained classification model. Therefore, we calculate the difference between the predicted classification for the perturbed and non-perturbed image as the score. Then we take the log of the inverse of that score as we look to minimize the loss score. 

\textbf{L1 Loss}
$ \frac{1}{n} \sum_{i=1}^{n}  |MRI_{p} - MRI_{np}|$
Where $MRI_{p}$ is the perturbation applied MRI volume and $MRI_{np}$ is the original MRI volume. 
After some experimentation, we observed that perturbation loss alone was not able to generate localized perturbation regions. To ensure we generated images that are more than random noise, we incorporated the L1 loss. L1 loss or Least Absolute Deviation loss minimizes the error sum of the given two images. This forces the model to learn to generate a perturbation map with the least amount of change to the original image, thus learning more localized regions. In the context of an adversely generative model, this acts as the minimizing element.

\textbf{Indecisive Penalty}
$\frac{1}{n} \sum_{i=1}^{n}  (-\alpha(y_{p}-\beta)^{2}+\delta)$
Where $\alpha,\beta$ and $\delta $ are hyperparameters for the parabolic function and $y_{p}$ is the probability of the perturbed input. After the initial experimentation, we also observed that the model training would plateau once it learns to predict at the center point (pretrained classification predicting 0.5). To penalize this behaviour, we introduced indecisive loss, which is a parabolic function with the highest penalty at the center point, to discourage the model from staying at the center point.

\textbf{Final Loss Function}
The final loss function is shown below.
\begin{center}
    \[\frac{1}{n} ( \sum_{i=1}^{n} log(\frac{1}{|y_{p} - y_{np}|})  +  \sum_{i=1}^{n}  -\alpha(y_{p}-\beta)^{2}+\delta) + \sum_{i=1}^{n}  |MRI_{p} - MRI_{np}|)\] 
\end{center}

\subsubsection{Generator}
We selected a modified U-net \cite{ronneberger2015u} architecture as our generator model. The model was modified to have a smaller bottleneck to reduce the likelihood of an identity function forming. The model was then trained with the input of 128 × 128 × 128 × 4 MRI volumes to output a mask with the same shape. Using the previously discussed loss function, the model was trained with a learning rate of 0.1 and Adam optimizer for 200 epochs. The best epoch was selected using the validation set for the proposed method’s evaluation.
\subsubsection{Classifier}
We have selected a 3D ResNet \cite{he2015deep} model as our classification architecture to evaluate the proposed method. The model was trained with a learning rate of 0.01 and the Adam optimizer \cite{kingma2014adam} for 100 epochs using binary cross-entropy as the loss function. The trained model at epoch 81 was selected for evaluation as it achieved the highest validation AUC of 0.86. The model achieved an area under the ROC curve of 0.83 on the test set.
\subsection{Relevance Map Generation}
 First, we generate the optimal perturbation mask for a given MRI scan. Then for each MRI sequence, we generate the superpixel maps with SLIC algorithm \cite{achanta2010slic}. Then for each superpixel, we assign the sum of values within the superpixel region in the generated perturbation mask. Finally, we combine all MRI sequences by summing each volume together into a single 3D volume. Then, we use binning to create ranked regions once the combined superpixel maps were generated. This allows us to pick top-ranking regions as the candidate ROI, which had the highest relevance to a given classification. 

\section{Experimentation}
For experimentation, the Multimodal Brain Tumour Segmentation Challenge 2020 dataset (BraTS) was used as our dataset \cite{menze2014multimodal,bakas2017advancing,bakas2018identifying,bakas2017segmentation}. 3D volumes of T1w, T1wCE, T2w and FLAIR sequences were available for each patient, along with the glioma tumour classification and segmentation. 190 3D scans (training : 133, validation : 19, test : 38 ) were used  to maintain a 60\% HGG and 40\% LGG tumour ratio. Each sequence was independently normalized using min-max normalization, and the center cropped to 128 $\times$ 128 $\times$ 128 volumes.

For evaluation, we computed the average Dice similarity coefficient (DSC) with optimal thresholding, which clusters multiple superpixels with the best grouping. Secondly, we calculated the average DSC for the highest-ranked superpixels based on the generated relevance ranking. Finally, we compare our results to the methods proposed by \cite{rajapaksa2021localized}, LIME \cite{ribeiro2016should} and Grad-CAM \cite{selvaraju2016grad}.  

\begin{figure*}
\begin{center}
\includegraphics[width=10cm]{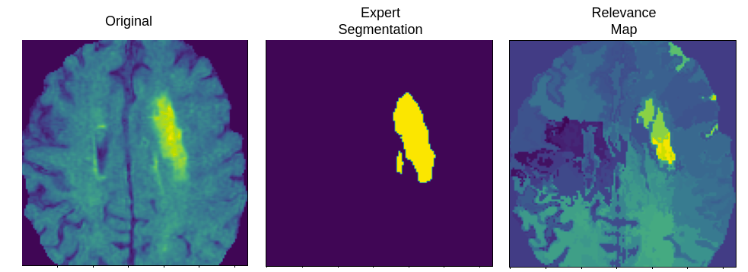}
\caption{Example of generated Relevance Map.}
\end{center}
\label{fig:overview}
\end{figure*}

\section{Results}

As shown in Table 1 our method outperformed all other methods with an average DSC score of 0.72 when using best superpixel clustering. As shown in Table 2. our approach also outperformed previously proposed Relevance Map methods when compared against each highly ranked superpixel. This indicates that our approach selected more of the tumour region than others on average. Figure 1 display an example Relevance Map slice. 

\begin{table}[]
\centering
\begin{tabular}{llllll}
\hline
                                                            & LIME \cite{ribeiro2016should} & Grad-CAM \cite{selvaraju2016grad} & \begin{tabular}[c]{@{}l@{}}Relevance Maps \\ (Blank \\ Perturbation) \cite{rajapaksa2021localized}\end{tabular} & \begin{tabular}[c]{@{}l@{}}Relevance Maps \\ ( Optimal Local\\  Perturbation \cite{rajapaksa2021localized})\end{tabular} & \begin{tabular}[c]{@{}l@{}} Optimal Global\\  Perturbation\\ (Ours)\end{tabular} \\ \hline
\begin{tabular}[c]{@{}l@{}}Average\\ DSC score\end{tabular} & 0.06 & 0.11     & 0.40                                                                              & 0.45                                                                                       & \textbf{0.72}                                                                             \\ \bottomrule
\end{tabular}

\caption{Average DSC score comparison between methods using best superpixel groupings.}

\end{table}

\begin{table}[t]
    \centering
   \begin{tabular}{llll}
\hline
\multicolumn{1}{l}{Rank} & \begin{tabular}[c]{@{}l@{}}Blank \\ Perturbation \cite{rajapaksa2021localized}\end{tabular} & \begin{tabular}[c]{@{}l@{}}Optimal \\ Local Perturbation \cite{rajapaksa2021localized}\end{tabular} & \begin{tabular}[c]{@{}l@{}} Optimal Global \\ Perturbation (Ours)  \end{tabular} \\ \hline
1                         & 0.25               & 0.31                         & 0.39                        \\ 
2                         & 0.07               & 0.18                         & 0.24                        \\ 
3                         & 0.06               & 0.10                         & 0.21                        \\  \bottomrule
\end{tabular}
\caption{Average DSC on ranked superpixels generated by different perturbation methods.}
 
    \label{tab:my_label}
\end{table}

\section{Acknowledgement}
This research has been supported by Huawei Technologies Canada Co., Ltd.

\bibliography{cite.bib}

\end{document}